\documentclass[12pt]{article}
\pdfoutput=1

\usepackage{graphicx,amssymb,amsmath,empheq}
\usepackage{amsthm,authblk}
\usepackage{empheq}
\usepackage{comment}
\usepackage{subfig} 
\usepackage[title]{appendix}
\usepackage{hyperref}
\hypersetup{colorlinks = true, linkcolor=black, citecolor=black, urlcolor=blue}
\usepackage{url}

\theoremstyle{plain}

\usepackage{tikz,fp}
\usepackage{tikz-cd}
\usetikzlibrary{arrows}
\usetikzlibrary{intersections}
\usetikzlibrary{shapes.geometric}
\usetikzlibrary{decorations.pathmorphing, patterns,shapes,fixedpointarithmetic}
\usetikzlibrary{decorations.markings}

\pgfdeclarepatternformonly{south west lines}{\pgfqpoint{-0pt}{-0pt}}{\pgfqpoint{3pt}{3pt}}{\pgfqpoint{3pt}{3pt}}{
        \pgfsetlinewidth{0.4pt}
        \pgfpathmoveto{\pgfqpoint{0pt}{0pt}}
        \pgfpathlineto{\pgfqpoint{3pt}{3pt}}
        \pgfpathmoveto{\pgfqpoint{2.8pt}{-.2pt}}
        \pgfpathlineto{\pgfqpoint{3.2pt}{.2pt}}
        \pgfpathmoveto{\pgfqpoint{-.2pt}{2.8pt}}
        \pgfpathlineto{\pgfqpoint{.2pt}{3.2pt}}
        \pgfusepath{stroke}}

\tikzset{
  mid arrow/.style={postaction={decorate,decoration={
        markings,
        mark=at position .575 with {\arrow{stealth}}
      }}},
  near arrow/.style={postaction={decorate,decoration={
        markings,
        mark=at position .275 with {\arrow{stealth}}
      }}},
  far arrow/.style={postaction={decorate,decoration={
        markings,
        mark=at position .800 with {\arrow{stealth}}
      }}},
  snake arrow/.style={fixed point arithmetic, decorate, decoration={snake,amplitude=2pt, segment length=11pt},postaction={decoration={markings,mark=at position 0.625 with {\arrow{stealth}}},decorate}},
}

\usepackage[nosort]{cite}

\title{More on Complex Sachdev-Ye-Kitaev Eternal Wormholes}

\author[1]{Pengfei Zhang}
\affil[1]{\normalsize\it California Institute of Technology, Pasadena, CA 91125, USA}

\begin{document}

\maketitle

\begin{abstract}
In this work, we study a generalization of the coupled Sachdev-Ye-Kitaev (SYK) model with $U(1)$ charge conservations. The model contains two copies of the complex SYK model at different chemical potentials, coupled by a direct hopping term. In the zero-temperature and small coupling limit with small averaged chemical potential, the ground state is an eternal wormhole connecting two sides, with a specific charge $Q=0$, which is equivalent to a thermofield double state. We derive the conformal Green's functions and determine corresponding IR parameters. At higher chemical potential, the system transit into the black hole phase. We further derive the Schwarzian effective action and study its quench dynamics. Finally, we compare numerical results with the analytical predictions.
\end{abstract}

\vspace{30pt}

\tableofcontents
\section{Introduction}
Proposed by Kitaev \cite{kitaev2014hidden} and related to the early work of Sachdev and Ye \cite{sachdev1993gapless}, the Sachdev-Ye-Kitaev (SYK) model \cite{maldacena2016remarks,kitaev2018soft} describes $N$ Majorana fermions with random $q$ body interactions. The SYK model can be solved in terms of the $1/N$ expansion. In the low-temperature limit, the model shows emergent conformal symmetry, and the scaling dimension of the Majorana operators is $1/q$. Such conformal symmetry is explicitly broken in the Lagrangian, which leads to a finite Schwarzian action for reparametrization modes \cite{maldacena2016remarks,kitaev2018soft}. Interestingly, such Schwarzian action matches the boundary action of the Jackiw-Teitelboim gravity in two-dimensional nearly anti-de Sitter (AdS2) spacetime, and consequently, the SYK model is a concrete model with low-energy holographic description. Later, the model is generalized to study various problems in both high energy and condensed matter communities \cite{maldacena2018eternal,sachdev2015bekenstein,davison2017thermoelectric,gu2020notes,Gnezdilov_2018,chen2017tunable,Kruchkov_2020,Altland_2019,can2019charge,banerjee2017solvable,chen2017competition,zhang2017dispersive,jian2017solvable,gu2017local,khveshchenko2019one,khveshchenko2020connecting,klebanov2020spontaneous,kim2019symmetry}. 

From the gravity perspective, a generalization by Maldancena and Qi \cite{maldacena2018eternal} is of special interest. They consider coupling two statistically correlated copies of the SYK model by a bilinear term, which is similar to the early proposal of traversable wormholes by Gao, Jafferis and Wall \cite{gao2017traversable}. Consequently, for small coupling strength, the ground state is dual to an eternal traversable wormhole in the global AdS$_2$ spacetime with $ds^2=(-dt^2+d\sigma^2)/\sin^2 \sigma$, where two copies of the SYK model lie on two boundaries near $\sigma=0$ and $\sigma=\pi$. Such a state can also be prepared as a thermofield double (TFD) state of the SYK model. At high temperatures, the coupled SYK model transits into the black hole phase, where two copies become disconnected geometrically. The finite temperature spectral of this coupled SYK model has been studied in \cite{Qi_2020,Plugge_2020}, and the real-time formation of the wormhole is discussed in \cite{maldacena2019syk}. Experimentally, the coupled SYK is found to be related to random spin models \cite{zhou2020disconnecting}, and there is also a proposal for realizing the MQ model \cite{lantagne2020diagnosing} in solid-state materials. The tunneling spectroscopy \cite{zhou2020tunneling} and the entanglement entropy \cite{chen2019entanglement} of the coupled SYK model has also been studied.

In this work, we study the effect of additional conservation law on the eternal traversable wormholes in the SYK-like models. To be concrete, we focus on the complex fermion version of the coupled SYK model with $U(1)$ symmetry. For a single SYK model, the generalization to complex fermions where the charge is conserved has been studied in details \cite{sachdev2015bekenstein,davison2017thermoelectric,gu2020notes}. There are also recent studies \cite{sahoo2020traversable,sorokhaibam2020traversable,garcia2020phase} on the complex version of the coupled SYK model. However, in these works, the chemical potential for two copies is fixed to be the same, which, as we will see, leads to the same wormhole state as no chemical potential term in the low-energy limit. We instead consider the case where the chemical potential of both systems can be tuned freely. In the low-energy limit, we find a family of conformal solutions for wormholes parameterized by three IR parameters $t'$, $\mathcal{P}$ and $\mathcal{E}$, which comes from the two chemical potentials and one coupling strength between two copies. At higher chemical potential, the conformal solution breaks down, and the system transits into black hole solutions. One can also derive the effective Schwarzian theory by requiring the diffeomorphism-invariance \cite{kitaev2018soft,gu2020notes}, where one could can the energy level of boundary gravitons and the quench dynamics. 

The paper is organized as follows: In section \ref{II}, we first describe our model and conformal solution, where the IR parameters can be fixed through the TFD State perspective, together with minimizing the effective action. In section \ref{III}, we compare our low-energy solutions to the numerics, in both imaginary-time and real-time approach. We finally summarize our results in \ref{IIII}.

\section{Model and its Low-energy Description} \label{II}
In this section, we present the generalized coupled SYK model with $U(1)$ symmetry and derive its low-energy behaviors. We consider two copies of the SYK$_4$ model with the same random interaction parameters:
\begin{equation}\label{Ham}
\begin{aligned}
H=&\frac{1}{4}\sum_{ijkl}J_{ijkl}c^\dagger_{L,i}c^\dagger_{L,j} c^{}_{L,k} c^{}_{L,l}+\frac{1}{4}\sum_{ijkl}J_{ijkl}c^\dagger_{R,i}c^\dagger_{R,j} c^{}_{R,k} c^{}_{R,l}\\
&-\mu_L \sum_i \left(c^\dagger_{L,i}c^{}_{L,i}-\frac{1}{2}\right)-\mu_R \sum_i \left(c^\dagger_{R,i}c^{}_{R,i}-\frac{1}{2}\right)-\mu_c \sum_i\left( c^\dagger_{R,i}c^{}_{L,i}+c^\dagger_{L,i}c^{}_{R,i}\right).
\end{aligned}
\end{equation}
Here the random coefficients $J_{ijkl}$ take distribution
\begin{equation}
\overline{J_{ijkl}}=0,\ \ \ \ \ \ \ \ \ \ \ \ \overline{|J_{ijkl}|^2}=2J^2/N^3.
\end{equation} 
We have added different chemical potential term $\mu_L/\mu_R$ for the left/right copy of the complex SYK model, with an additional directly hopping term $\mu_c$. Without loss of generality, we have set the hopping amplitude to be real. For later convenience, we introduce the averaged $\bar{\mu}$ and the relative chemical potential $\mu$ as
\begin{equation}\label{def}
\frac{\mu_L-\mu_R}{2}=\mu,\ \ \ \ \ \ \ \ \ \ \ \ \ \ \frac{\mu_L+\mu_R}{2}=\overline{\mu}.
\end{equation}

 Before analyzing the model, we add a few comments:
\begin{enumerate}
\item Although we introduced the model with two copies of the complex SYK$_4$ system for simplicity, it is straightforward to generalize all discussions in the paper for other even finite $q\leq 4$. Generally, the random interaction strength of the right copy should contain an additional factor of $(-1)^{q/2}$, as in the Majorana case \cite{maldacena2018eternal}. For non-zero chemical potential, the large-$q$ limit becomes problematic, similar to the single SYK case \footnote{We thank Yingfei Gu for the discussion of the large-$q$ limit of a single complex SYK model.}.

\item Without the hopping term, the model contains two copies of the complex SYK model, and the symmetry is $U(1)\times U(1)$, which is broken to a single $U(1)$ by $\mu_c$. The idea why here we still introduce two different chemical potential terms is that for eternal wormholes, we are focusing on the limit of small $\mu_c \rightarrow 0$. In this limit, the breaking of the relative phase transformation $c^{}_{L,i}\rightarrow c^{}_{L,i}e^{i\alpha},\ c^{}_{R,i}\rightarrow c^{}_{R,i}e^{-i\alpha}$ is weak and there exists a corresponding pseudo Goldstone mode. Therefore we expect a relative chemical potential can play important roles in the low-energy limit. 

\item Such a model can be understood as certain spinful SYK model or large-$M$ spin models \cite{zhou2020disconnecting}, where $L/R$ can be understood as different spin components $\uparrow$ and $\downarrow$. In such models, the relative chemical potential $\mu$ corresponds to a magnetic field in the $z$ direction, while the coupling term $\mu_c$ represents a magnetic field in the $x$ direction. By a particle-hole mapping, this can also be understood as a spinful SYK model with a proximity pairing term.

\end{enumerate}
\subsection{Conformal Solution}
We consider the model in the imaginary-time formalism. We focus on the large-$N$ limit. Under the replica diagonal assumption, after introducing the bilocal fields $G(\tau,\tau')$, $\Sigma(\tau,\tau')$ and integrating out fermions, the partition function becomes:
\begin{equation}\label{Action}
\begin{aligned}
\frac{S[G,\Sigma]}{N}=&-\text{Tr}\left[\log(\sigma_{\alpha\beta}-\Sigma_{\alpha\beta})\right]-\int d\tau d\tau'\left[G_{\alpha\beta}(\tau,\tau')\Sigma_{\beta\alpha}(\tau',\tau)+\frac{J^2}{4}G_{\alpha\beta}^2(\tau,\tau')G_{\beta\alpha}^2(\tau',\tau)\right]\\
=&-\text{Tr}\left[\log(-\tilde{\Sigma}_{\alpha\beta})\right]-\int d\tau d\tau'\left[G_{\alpha\beta}(\tau,\tau')\tilde{\Sigma}_{\beta\alpha}(\tau',\tau)+\frac{J^2}{4}G_{\alpha\beta}^2(\tau,\tau')G_{\beta\alpha}^2(\tau',\tau)\right]\\
&-\int d\tau d\tau'\ G_{\alpha\beta}(\tau,\tau')\sigma_{\beta\alpha}(\tau',\tau).
\end{aligned}
\end{equation}
Here $\alpha,\beta=L,R$ labels two copies of the SYK model, $G_{\alpha\beta}(\tau,\tau')=\sum_i c^{}_{\alpha,i}(\tau)\bar{c}_{\beta,i}(\tau')/N$ is the fluctuation of fermion bilinear. We have defined $\Sigma=\tilde{\Sigma}+\sigma$, with the UV perturbation \cite{kitaev2018soft,gu2020notes}
\begin{equation}
\sigma(\tau,\tau')=\begin{pmatrix}
\delta'(\tau-\tau')-\mu_L\delta(\tau-\tau')&-\mu_c\delta(\tau-\tau')\\
-\mu_c\delta(\tau-\tau')&\delta'(\tau-\tau')-\mu_R\delta(\tau-\tau')
\end{pmatrix}.
\end{equation}
Without $\sigma$, there is an emergent conformal symmetry of bilocal fields:
\begin{equation}\label{csymmetry}
\begin{aligned}
G_{\alpha\beta}(\tau,\tau')&\rightarrow \varphi_\alpha'(\tau)^{\frac{1}{4}}\varphi_\beta'(\tau')^{\frac{1}{4}}e^{i\lambda_\alpha(\tau)-i\lambda_\beta(\tau')}G_{\alpha\beta}\left(\varphi_\alpha(\tau),\varphi_\beta(\tau')\right),\\
\Sigma_{\alpha\beta}(\tau,\tau')&\rightarrow \varphi_\alpha'(\tau)^{\frac{3}{4}}\varphi_\beta'(\tau')^{\frac{3}{4}}e^{i\lambda_\alpha(\tau)-i\lambda_\beta(\tau')}\Sigma_{\alpha\beta}\left(\varphi_\alpha(\tau),\varphi_\beta(\tau')\right),
\end{aligned}
\end{equation}
the consequence of this emergent symmetry will be studied later.

Now, we take the saddle-point equations of \eqref{Action} for bilocal fields. It corresponds to the Schwinger-Dyson equation of fermions: 
\begin{equation}\label{SD}
\left(G^{-1}(\omega)\right)_{\alpha\beta}= -\tilde{\Sigma}_{\alpha\beta}(\omega) =\sigma_{\alpha\beta}(\omega)-\Sigma_{\alpha\beta}(\omega),\ \ \ \ \ \ \Sigma_{\alpha\beta}(\tau)=-J^2G_{\alpha\beta}(\tau)^2G_{\beta\alpha}(-\tau).
\end{equation}
In the low-energy limit $J|\tau|\gg 1$, we hope to neglect $\sigma$ and consider the zero-temperature solution of 
\begin{equation}\label{SDconformal}
\sum_\gamma \int_{-\infty}^{\infty} d\tau'' G_{\alpha\gamma}(\tau-\tau'')J^2G_{\gamma\beta}(\tau''-\tau')^2 G_{\beta\gamma}(\tau'-\tau'') =\delta_{\alpha\beta}\delta(\tau-\tau').
\end{equation}
The validity of \eqref{SDconformal} needs further explanations: When $\mu_c=0$ and $\bar{\mu}=0$, the system represents decoupled complex SYK models with opposite chemical potentials $\pm \mu$ with the scaling dimension of fermion operators $1/4$. Consequently, the perturbation from left-right coupling is relevant and the system should be gapped. Consequently, to neglect the UV symmetry breaking term $\sigma$, we should require the solution of \eqref{SDconformal} leads to a gap much larger than the bare gap $\mu_c$. As we will see later, this requires $\bar{\mu}/J\ll1$ and $\mu_c/J\ll1$ \cite{maldacena2018eternal}, while $\mu/J$ can be $O(1)$.

Interestingly, there exists a family of conformal solutions with three IR parameters $\mathcal{E}$, $\mathcal{P}$ and $t'$. This is what we expect since there are three UV parameters $\mu_L$, $\mu_R$, and $\mu_c$ in $\sigma$. Explicitly, we have
\begin{equation}\label{conformal}
\begin{aligned}
G_{LL}(\tau)&=b\times \text{sgn}(\tau)e^{\pi\mathcal{E}\text{sgn}(\tau)+\mathcal{P}\tau}\left(\frac{t'}{2\sinh (t' |\tau|/2)}\right)^{1/2},\\
G_{RR}(\tau)&=b\times \text{sgn}(\tau)e^{-\pi\mathcal{E}\text{sgn}(\tau)+\mathcal{P}\tau}\left(\frac{t'}{2\sinh (t' |\tau|/2)}\right)^{1/2},\\
G_{LR}(\tau)&=G_{RL}(\tau)=-b\times e^{\mathcal{P}\tau}\left(\frac{t'}{2\cosh (t' |\tau|/2)}\right)^{1/2}.
\end{aligned}
\end{equation}
Here the overall coefficient $b$ satisfies $4\pi J^2 \cosh(2\pi\mathcal{E})b^4=1$. $G_{LR}(\tau)=G_{RL}(\tau)$ is a consequence of the time-reversal symmetry. To make the Green's function well-defined as $|\tau| \rightarrow \infty$, we require that $\mathcal{P}<t'/4$. In the following discussion, we will always neglect terms $\sim t'^2$. For $|\tau|t'\ll1$, $G_{LL}$ and $G_{RR}$ reduce to the zero-temperature Green's function of decoupled complex SYK model with asymmetry parameters $\pm\mathcal{E}$ \cite{davison2017thermoelectric,gu2020notes}:
\begin{equation}\label{single}
G_{LL}(\tau) = -G_{RR}(-\tau)= be^{\pi\mathcal{E}\text{sgn}(\tau)}\frac{\text{sgn}(\tau)}{|\tau|^{1/2}}.
\end{equation}
On the contrary, when both $\mu=0$ and $\bar{\mu}=0$, we should get back to the Majorana wormhole solution \cite{maldacena2018eternal}, which corresponds to $\mathcal{E}=\mathcal{P}=0$. When we turn on $\mu$ with $\bar{\mu}=0$, the ensemble of the system is still invariant under a particle-hole transformation $c^{}_{L,i} \rightarrow c^\dagger_{R,i}$. This is compatible with having a non-vanishing $\mathcal{E}$. We also have the gap of the fermionic excitations $E_g=t'/4$. On the other hand, the existence of $\mathcal{P}$ breaks this particle-hole symmetry, which corresponds to the existence of $\bar{\mu}$. The relation between the IR parameters and the UV parameters or physical observables are discussed in the next few subsections, while the results are summarized in the Table \ref{tab}.
\begin{table}[t]
  \centering
\begin{tabular}{|c|c|c|c|}
\hline
 IR parameters& $t'$ & $\mathcal{P}$ & $\mathcal{E}$ \\ \hline
 Relations& $t'=\frac{b^{2/3} J^{2/3} \mu _c^{2/3}}{2 \alpha _S^{2/3}}$ & $\mathcal{P}=\bar{\mu}$ & $Q_L=-\frac{\theta}{\pi}-\frac{1}{4}\sin(2\theta),\ e^{2\pi \mathcal{E}}=\tan (\pi/4+\theta)$ \\ \hline
\end{tabular}
\caption{A summary of relations between the IR parameters and UV parameters/physical observables in the eternal wormhole phase.}
\label{tab}
\end{table}

In frequency space, \eqref{conformal} becomes
\begin{equation}\label{conformalw}
\begin{aligned}
G_{LL}(\omega+i\mathcal{P})&=b\times\frac{2 \sqrt{2} \pi ^{3/2} \text{sech}(2 \pi  \omega/t' ) (i \sinh (\pi  \omega/t' ) \cosh (\pi  \mathcal{E} )+\cosh (\pi  \omega/t') \sinh (\pi
    \mathcal{E} ))}{\Gamma
   \left(\frac{3}{4}-i \frac{\omega}{t'} \right) \Gamma \left(i \frac{\omega}{t'} +\frac{3}{4}\right)},\\
  G_{RR}(\omega+i\mathcal{P})&=b\times\frac{2 \sqrt{2} \pi ^{3/2} \text{sech}(2 \pi  \omega/t' ) (i \sinh (\pi  \omega/t' ) \cosh (\pi  \mathcal{E} )-\cosh (\pi  \omega/t') \sinh (\pi
    \mathcal{E} ))}{\Gamma
   \left(\frac{3}{4}-i \frac{\omega}{t'} \right) \Gamma \left(i \frac{\omega}{t'} +\frac{3}{4}\right)},\\
  G_{LR}(\omega+i\mathcal{P})&= G_{RL}(\omega+i\mathcal{P})=-b\times\frac{\Gamma \left(\frac{1}{4}\right) \Gamma \left(\frac{3}{4}\right) \Gamma
   \left(\frac{1}{4}-i \frac{\omega}{t'} \right) \Gamma \left(i \frac{\omega}{t'} +\frac{1}{4}\right)}{\sqrt{2t'}\pi
   ^{3/2}}.
   \end{aligned}
\end{equation}
Here the $G_{\alpha\beta}$ is analytic on the stripe with $|\text{Im}(\omega+i\mathcal{P})|<t'/4$, as anticipated from the real-time expressions. There is a similar expression for the self energy $\Sigma_{\alpha\beta}$, and it is straightforward to check they satisfies the Schwinger-Dyson equation \eqref{SDconformal}. 

Here we present another method which allow us to derive \eqref{conformal} from the single SYK conformal solution \eqref{single}. For conciseness, let us introduce $G_0(\tau,\tau')=b\ {\text{sgn}(\tau-\tau')}/{|\tau-\tau'|^{1/2}}$ and $\Sigma_0(\tau,\tau')=b^3\ {\text{sgn}(\tau-\tau')}/{|\tau-\tau'|^{3/2}}$. It is known that \eqref{single} satisfies the self-consistent equation \cite{davison2017thermoelectric}:
\begin{equation}
-J^2\int^\infty_{-\infty} dt_3\ e^{\pi\mathcal{E}\left(\text{sgn}(t_{13})+\text{sgn}(t_{32})\right)}G_0(t_1,t_3) \Sigma_0(t_3,t_2)=\delta(t_1-t_2).
\end{equation}
Here $\tau_{ij}=\tau_{i}-\tau_{j}$. Now we split the integral of $t_3$ into $(-\infty,0)$ and $(0,\infty)$, which gives
\begin{equation}
-J^2\left(\int^0_{-\infty} dt_3+\int^\infty_{0} dt_3\right) e^{\pi\mathcal{E}\left(\text{sgn}(t_{13})+\text{sgn}(t_{32})\right)}G_0(t_1,t_3) \Sigma_0(t_3,t_2)=\delta(t_1-t_2).
\end{equation}
As an example, we consider $t_1>0$ and $t_2>0$. We define new time variable $\tau$ as 
\begin{equation}
t=\left\{
\begin{aligned}
e^{t'\tau} & \ \ \ \  t>0, \\
-e^{t'\tau} & \ \ \ \  t<0.
\end{aligned}
\right.
\end{equation}
Here $\tau \in (-\infty,\infty)$. This change of variable can be motivated from the gravity picture, where one can map the Poincare plane to the gAdS$_2$ stripe. This gives 
\begin{equation}
\begin{aligned}
\delta(\tau_1-\tau_2)&\left|\frac{d\tau_1}{dt_1}\right|^{1/4}\left|\frac{d\tau_2}{dt_2}\right|^{3/4}=\\
&-J^2\int^\infty_{-\infty} d\tau_3 \left|\frac{dt_3}{d\tau_3}\right|e^{\pi\mathcal{E}\text{sgn}(\tau_{13})}G_0(e^{t'\tau_1},e^{t'\tau_3}) e^{\pi\mathcal{E}\text{sgn}(\tau_{32})}\Sigma_0(e^{t'\tau_3},e^{t'\tau_2})\\
&-J^2\int^\infty_{-\infty} d\tau_3 \left|\frac{dt_3}{d\tau_3}\right|G_0(e^{t'\tau_1},-e^{t'\tau_3}) \Sigma_0(-e^{t'\tau_3},e^{t'\tau_2}).
\end{aligned}
\end{equation}
Inversing the factors on the L.H.S., we find 
\begin{equation}
\begin{aligned}
\delta&(\tau_1-\tau_2)=\\
&-J^2\int^\infty_{-\infty} d\tau_3 \underbrace{e^{\pi\mathcal{E}\text{sgn}(\tau_{13})+\mathcal{P}\tau_{13}}\left|\frac{dt_1}{d\tau_1}\frac{dt_3}{d\tau_3}\right|^{1/4}G_0(e^{t'\tau_1},e^{t'\tau_3})}_{G_{LL}} \underbrace{e^{\pi\mathcal{E}\text{sgn}(\tau_{32})+\mathcal{P}\tau_{32}}\left|\frac{dt_3}{d\tau_3}\frac{dt_2}{d\tau_2}\right|^{3/4}\Sigma_0(e^{t'\tau_3},e^{t'\tau_2})}_{\Sigma_{LL}}\\
&-J^2\int^\infty_{-\infty} d\tau_3  \underbrace{e^{\mathcal{P}\tau_{13}}\left|\frac{dt_1}{d\tau_1}\frac{dt_3}{d\tau_3}\right|^{1/4}G_0(e^{t'\tau_1},-e^{t'\tau_3})}_{-G_{LR}}\underbrace{e^{\mathcal{P}\tau_{32}}\left|\frac{dt_3}{d\tau_3}\frac{dt_2}{d\tau_2}\right|^{3/4} \Sigma_0(-e^{t'\tau_3},e^{t'\tau_2})}_{-\Sigma_{RL}},
\end{aligned}
\end{equation}
where we have used the (imaginary version of) $U(1)$ conformal symmetry \eqref{csymmetry} to introduce $\mathcal{P}$, and identified the $G_{\alpha\beta}$ and $\Sigma_{\alpha\beta}$, which match the definition \eqref{conformal}. This equation is exactly \eqref{SDconformal} with $\alpha=\beta=L$. Other components correspond to taking $t_1$ and $t_2$ in different regions ($t_1\lessgtr 0$ and $t_2\lessgtr 0$).

With the conformal solution, we could also determine the conserved $U(1)$ charge, as for a single complex SYK model \cite{gu2020notes}. We define the charge per mode $Q$ as 
\begin{equation}
Q=Q_L+Q_R=\left<c^\dagger_{L,i}c^{}_{L,i}-\frac{1}{2}\right>+\left<c^\dagger_{R,i}c^{}_{R,i}-\frac{1}{2}\right>=-G_{LL}(0^-)-G_{RR}(0^-)-1.
\end{equation}
Using the fact that $G_{\alpha\alpha}(0^+)-G_{\alpha\alpha}(0^-)=1$, we have
\begin{equation}
Q=-\frac{\text{tr}[G(0^+)+G(0^-)]}{2}=\int d\tau\ \text{tr}\left[\sigma(\tau)G(-\tau)\right]\tau=-\int^\infty_{-\infty} d\tau\ \text{tr}\left[\tilde{\Sigma}(\tau)G(-\tau)\right]\tau.
\end{equation}
This can also be derived by applying Noether's theorem of the $U(1)$ symmetry for the action \eqref{Action}. Note that different from the single complex SYK case, the integral converges absolutely in the long time limit, and there is no need to introduce additional regulators. 

To work out the result, we change to the frequency domain:
\begin{equation}
Q=-\frac{1}{2\pi i}\int^\infty_{-\infty} d\omega\ \partial_\omega\text{tr}\left[\log G(\omega)\right]=-\frac{1}{2\pi i}\left(\int_{|\omega|<\Lambda}d\omega+\int_{|\omega|>\Lambda}d\omega\right)\partial_\omega\text{tr}\left[\log G(\omega)\right],
\end{equation} 
where we cut the integral into two parts with $\Lambda\sim J\gg t'$. In the second part of the integral, we are considering the frequency regime where the Green's function takes the same form as a single complex SYK model. Consequently, the result matches the corresponding calculation of the single complex model \cite{gu2020notes}:
\begin{equation}
-\frac{1}{2\pi i}\int_{|\omega|>\Lambda}d\omega\ \partial_\omega\text{tr}\left[\log G(\omega)\right]=\frac{1}{2\pi i}(-2i\theta+2i\theta)=0,
\end{equation}
where $\theta$ is related to $\mathcal{E}$ as $e^{2\pi \mathcal{E}}=\tan (\pi/4+\theta)$. For the integral over small frequency, we could directly use conformal solutions \eqref{conformalw}:
\begin{equation}
-\frac{1}{2\pi i}\int_{-\Lambda}^{\Lambda}d\omega\ \partial_\omega\text{tr}\left[\log G(\omega)\right]=-\frac{1}{2\pi i}\int_{-\Lambda}^{\Lambda}dz\ \partial_z\text{tr}\left[\log G(z+i\mathcal{P})\right]+O((\mathcal{P}/J)^2)\approx 0.
\end{equation}
We have deformed the contour to absorb the presence of $\mathcal{P}$, and after the deformation, the integral becomes zero. We also neglect high-order terms in $\mathcal{P}/J$ (or $t'/J$). Finally, we find the specific charge $Q=0$ for any valid wormhole solution \eqref{conformal}. This is consistent with the TFD perspective discussed in the next subsection.

\subsection{the TFD State Viewpoint}\label{TFD sec}
Before the introduction of the SYK model, there is already a study on the consequence of coupling two 1+1-D CFTs with different chiralities \cite{qi2012general}, where the ground state is argued to be a TFD state. In \cite{maldacena2018eternal}, authors find the coupled SYK model show similar behaviors with the ground state being the TFD state. From the gravity perspective, this is dual to the vacuum of the gAdS$_2$ spacetime can be prepared by the path-integral over half of a disk.

In this subsection, we assume similar ideas hold for the complex version as well. A justification will be given in the next section. To construct the TFD state, we firstly consider the ground state of the relevant hopping term $-\mu_c \sum_i\left( c^\dagger_{R,i}c^{}_{L,i}+c^\dagger_{L,i}c^{}_{R,i}\right)$, which is simply the maximally entangled state
\begin{equation}
|I\rangle=\otimes_i \frac{1}{\sqrt{2}}\left(|0\rangle_{L,i} |1\rangle_{R,i}+|1\rangle_{L,i} |0\rangle_{R,i} \right),
\end{equation} 
with all symmetric modes $c^{}_{L,i}+c^{}_{R,i}$ being occupied. The state satisfies
\begin{equation}
c^{}_{L,i}|I\rangle=c^{}_{R,i}|I\rangle,\ \ \ \ \ \ \ \ \ c^{\dagger}_{L,i}|I\rangle=-c^{\dagger}_{R,i}|I\rangle.
\end{equation}

To take other terms in the Hamiltonian into account, we assume that to the leading order of small $\mu_c/J$, the ground state of \eqref{Ham} is a TFD state with an effective Hamiltonian $H_\text{eff}$:
\begin{equation}
|G\rangle=\frac{1}{\sqrt{Z}}\exp(-\beta_{\text{eff}}H_\text{eff}/4)|I\rangle,
\end{equation}
where $Z_{\text{eff}}$ is the normalization factor and $H_\text{eff}$ is given by
\begin{equation}
\begin{aligned}
H_\text{eff}=&\frac{1}{4}\sum_{ijkl}J_{ijkl}c^\dagger_{L,i}c^\dagger_{L,j} c^{}_{L,k} c^{}_{L,l}+\frac{1}{4}\sum_{ijkl}J_{ijkl}c^\dagger_{R,i}c^\dagger_{R,j} c^{}_{R,k} c^{}_{R,l}\\
&-\mu_L'\sum_i c^\dagger_{L,i}c^{}_{L,i}-\mu_R' \sum_i c^\dagger_{R,i}c^{}_{R,i}.
\end{aligned}
\end{equation}
In general, we should allow $(\mu_L',\mu_R')$ being different from $(\mu_L,\mu_R)$. We can then define $\mu'$ and $\bar{\mu}'$ similar to \eqref{def}. Since we are focusing on the small $t'$ and thus small $\bar{\mu}$ limit, we expect $\bar{\mu}'\propto \bar{\mu}$ since they both correspond to the particle-hole symmetry breaking. Moreover, since both $\mu$ and $\beta_{\text{eff}}$ is particle-hole symmetric, we expect their $\bar{\mu}$ dependence is at least second-order, which can be safely neglected. For $\mu_c=0$, we expect $|G\rangle$ to be a specific pure state within the ground state sector of $H$, which is two decoupled complex SYK models. Consequently, we expect $\beta_{\text{eff}}\rightarrow \infty$ and $\mu'=\mu$ for $\mu_c\rightarrow 0$, if the coupled ground state is in the same charge sector as the decoupled one. The precise relations are given in Table \ref{tab2}, which will be proved in the next two subsections. 

\begin{table}[t]
  \centering
\begin{tabular}{|c|c|c|c|}
\hline
TFD parameters& $\beta_{\text{eff}}$ & $\bar{\mu}'$ & $\mu'$ \\ \hline
 Relations& $t' \beta_{\text{eff}}=2\pi$ & Redundant & $\mu-\mu'=t'\mathcal{E}$ \\ \hline
\end{tabular}
\caption{A summary of relations between the TFD parameters, IR parameters and parameters in the Hamiltonian \eqref{Ham}.}
\label{tab2}
\end{table}

We can directly use $|G\rangle$ to determine $\mathcal{E}$ and $\mathcal{P}$. We first notice that 
\begin{equation}
c^\dagger_{L,i}c^{}_{L,i}|I\rangle=(1-c^\dagger_{R,i}c^{}_{R,i})|I\rangle.
\end{equation}
Consequently, we have 
\begin{equation}
\left(c^\dagger_{L,i}c^{}_{L,i}+c^\dagger_{R,i}c^{}_{R,i}\right)|G\rangle=|G\rangle.
\end{equation}
As a result, we know $Q=0$, consistent with the result in the last subsection, and the $\overline{\mu}'$ term, and thus the $\overline{\mu}$ term, would not change the ground state $|G(\bar{\mu})\rangle=|G(0)\rangle$. As a result, we expect the only effect of $\overline{\mu}$ in terms of the Green's function is:
\begin{equation}
\begin{aligned}
G_{\alpha \beta }(\tau)&=\left<G \right|c^{}_{\alpha,i}(\tau)c^{\dagger}_{\beta,i}\left|G\right>=\left<G \right|e^{\tau H(\overline{\mu})}c^{}_{\alpha,i}e^{-\tau H(\overline{\mu})}c^{\dagger}_{\beta,i}\left|G\right>\\
&=e^{\overline{\mu}\tau}\left<G \right|e^{\tau H(0)}c^{}_{\alpha,i}e^{-\tau H(0)}c^{\dagger}_{\beta,i}\left|G\right>=e^{\overline{\mu}\tau}G_{\alpha\beta}(\tau)_{\overline{\mu}=0}
\end{aligned}
\end{equation}
This suggests that we could directly identify $\mathcal{P}=\overline{\mu}$.

We then consider the reduced density matrix of a single side of the system. Noticing that 
\begin{equation}
\begin{aligned}
 H_\text{eff}|\text{EPR}\rangle=& \left(\frac{1}{2}\sum_{ijkl}J_{ijkl}c^\dagger_{L,i}c^\dagger_{L,j} c^{}_{L,k} c^{}_{L,l}-2\mu' \sum_i c^\dagger_{L,i}c^{}_{L,i}-\mu_R'N\right)|\text{EPR}\rangle \equiv 2H_L|\text{EPR}\rangle\\
=&\left(\frac{1}{2}\sum_{ijkl}J_{ijkl}c^\dagger_{R,i}c^\dagger_{R,j} c^{}_{R,k} c^{}_{R,l}+2\mu' \sum_i c^\dagger_{R,i}c^{}_{R,i}-\mu_L'N\right)|\text{EPR}\rangle\equiv 2H_R|\text{EPR}\rangle.
\end{aligned}
\end{equation}
Consequently, if we look at a single side, the system is in the thermal ensemble at inverse temperature $\beta_{\text{eff}}$ with chemical potential $\pm \mu'$. When we evolve the system with $H$, the Greens function $G_{LL}$ takes the form of \eqref{conformal}. On the other hand, we could consider evolving the system with only $H_0$, which does not include the left-right coupling term in \eqref{Ham}, this leads to
\begin{equation}\label{quench}
\begin{aligned}
\tilde{G}_{LL}(\tau)&=\left<G \right|e^{\tau H_0}c^{}_{L,i}e^{-\tau H_0}c^{\dagger}_{L,i}\left|G\right>=e^{\overline{\mu}\tau+\delta \mu\tau}\left<G \right|e^{\tau H_L}c^{}_{L,i}e^{-\tau H_L}c^{\dagger}_{L,i}\left|G\right>=e^{\overline{\mu}\tau+\delta \mu\tau} G_{\beta_{\text{eff},\mathcal{E}_s(\mu')}}(\tau),
\end{aligned}
\end{equation}
Where $\delta \mu=\mu-{\mu}'$ is from the difference between $H_{\text{eff}}$ and $H_0$. Here the subscript in $\mathcal{E}_s(\mu')$ reminds us that it is defined for a single complex SYK model. $G_{\beta,\mathcal{E}}(\tau)$ is the finite temperature Green's function for a single complex SYK model:
\begin{equation}
G_{\beta,\mathcal{E}}(\tau)=b\times e^{2\pi\mathcal{E}(1/2-\tau/\beta)}\left(\frac{\pi}{\beta\sin (\pi\tau/\beta)}\right)^{1/2},\ \ \ \ \ \text{for}\ 0<\tau<
\beta.
\end{equation}
Since we focus on the limit where $\mu_c \ll J$ which results in $t'\ll J$, we expect there is a separation of time scale: when $1/J\ll\tau \ll 1/t'$, there should be no difference between $G_{LL}(\tau)$ and $\tilde{G}_{LL}(\tau)$. Matching the expression in this limit, we find their asymmetric factors should be identified $\mathcal{E}=\mathcal{E}_s(\mu')$. This function between $\mu'$ and $\mathcal{E}$ contains the UV data and no analytical formula is available, while $\mathcal{E}$ can be related to the charge \cite{davison2017thermoelectric,gu2020notes}: 
\begin{equation}
Q_L=-\frac{\theta}{\pi}-\frac{1}{4}\sin(2\theta).
\end{equation}
Finally, the entanglement entropy between two sides should match the thermal entropy 
\begin{equation}\label{entropy}
\mathcal{S}(\rho_L)=\mathcal{S}_0(Q_L)+4\pi^2 \frac{\alpha_S }{\beta_{\text{eff}}J},
\end{equation}
where $\mathcal{S}_0(Q)$ is the zero-temperature entropy of a single complex SYK model with charge $Q$ \cite{gu2020notes}.

\subsection{Effective Action and Thermodynamics}\label{effective sec}
Now we turn to the derivation of the effective action. Using the idea in \cite{kitaev2018soft,gu2020notes}, the effective action describes the coupling between the UV source $\sigma$ and IR fluctuations, which corresponds to the last term in \eqref{Action}. To derive the effective action, we first determine the IR fields. Separating out the $U(1)$ and reparametrization modes, we consider the quasi-solutions (for $\tau_1>\tau_2$):
\begin{equation}\label{IRqusi}
\begin{aligned}
G_{LL}(\tau_1,\tau_2)&=b\times e^{\pi\mathcal{E}+\mathcal{P}(\tau_1-\tau_2)}\left(\frac{\varphi_L'(\tau_1)\varphi_L'(\tau_2)}{4\sinh^2 (\varphi_L(\tau_1)-\varphi_L(\tau_2)/2)}\right)^{1/4}e^{i\left(\tilde{\lambda}_L(\tau_1)-\tilde{\lambda}(\tau_2)\right)},\\
G_{RR}(\tau_1,\tau_2)&=b\times e^{-\pi\mathcal{E}+\mathcal{P}(\tau_1-\tau_2)}\left(\frac{\varphi_R'(\tau_1)\varphi_R'(\tau_2)}{4\sinh^2 (\varphi_R(\tau_1)-\varphi_R(\tau_2)/2)}\right)^{1/4}e^{i\left(\tilde{\lambda}_R(\tau_1)-\tilde{\lambda}_R(\tau_2)\right)},\\
G_{LR}(\tau_1,\tau_2)&=-b\times e^{\mathcal{P}(\tau_1-\tau_2)}\left(\frac{\varphi_L'(\tau_1)\varphi_R'(\tau_2)}{4\cosh^2 (\varphi_L(\tau_1)-\varphi_R(\tau_2)/2)}\right)^{1/4}e^{i\left(\tilde{\lambda}_L(\tau_1)-\tilde{\lambda}_R(\tau_2)\right)},\\
G_{RL}(\tau_1,\tau_2)&=-b\times e^{\mathcal{P}(\tau_1-\tau_2)}\left(\frac{\varphi_R'(\tau_1)\varphi_L'(\tau_2)}{4\cosh^2 (\varphi_R(\tau_1)-\varphi_L(\tau_2)/2)}\right)^{1/4}e^{i\left(\tilde{\lambda}_R(\tau_1)-\tilde{\lambda}_L(\tau_2)\right)},
\end{aligned}
\end{equation}
where we have defined $\tilde{\lambda}_\alpha(\tau)-i\mathcal{P}\tau=\lambda_\alpha(\tau)-i\mathcal{P}\varphi_\alpha(\tau)$. 

We first consider the diagonal terms in $\sigma$, which are coupled to $G_{LL}$ and $G_{RR}$. The IR fields can be identified by expanding the quasi-solution around the intermediate asymptotic expression for $1/J \ll |\tau_1-\tau_2| \ll 1/t'$. We have:
\begin{equation}
\begin{aligned}
G_{LL}(\tau_1,\tau_2)&=G_{\infty,\mathcal{E}}(\tau_1-\tau_2)(1+A_L(\tau_+)(\tau_1-\tau_2)+B_L(\tau_+)(\tau_1-\tau_2)^2+...),\\
G_{RR}(\tau_1,\tau_2)&=G_{\infty,-\mathcal{E}}(\tau_1-\tau_2)(1+A_R(\tau_+)(\tau_1-\tau_2)+B_R(\tau_+)(\tau_1-\tau_2)^2+...).
\end{aligned}
\end{equation}
with $\tau_+=\frac{\tau_1+\tau_2}{2}$ and
\begin{equation}
A_\alpha(\tau)=i\tilde{\lambda}_\alpha'(\tau)+\mathcal{P},\ \ \ \ \ \ \ \ \ \ \ B_\alpha(\tau)=\frac{1}{24}\text{Sch}\left(\tanh\frac{\varphi_\alpha(\tau)}{2},\tau\right) +\frac{A_\alpha(\tau)^2}{2}.
\end{equation}
Here the Schwarzian derivative is defined as $\text{Sch}\left(y(\tau),\tau\right)\equiv \frac{y'''}{y'}-\frac{3}{2}\left(\frac{y''}{y'}\right)^2$. The general diffeomorphism and gauge invariance \cite{gu2020notes} suggest the first contribution to the effective action takes the form
\begin{equation}
\begin{aligned}
\frac{S^{(1)}_{\text{eff}}}{N}=&-\frac{\alpha_S}{J}\int d\tau \left[\text{Sch}\left(\tanh\frac{\varphi_L(\tau)}{2},\tau\right)+\text{Sch}\left(\tanh\frac{\varphi_R(\tau)}{2},\tau\right)\right]\\
&+\int d\tau \left[f(\mu_L-A_L)+f(\mu_R-A_R)\right]\\
=&-\frac{\alpha_S}{J}\int d\tau \left[\text{Sch}\left(\tanh\frac{\varphi_L(\tau)}{2},\tau\right)+\text{Sch}\left(\tanh\frac{\varphi_R(\tau)}{2},\tau\right)\right]\\
&+\int d\tau \left[f\left(\mu-i\tilde{\lambda}_L'(\tau)\right)+f\left(-\mu-i\tilde{\lambda}_R'(\tau)\right)\right].
\end{aligned}
\end{equation}
Here we have used the fact that $\mathcal{P}=\bar{\mu}$. The result has no dependence on $\bar{\mu}$, and this implies $Q=0$, making the substitution of $\mathcal{P}$ consistent. The function $f(\mu)$ represents the UV contribution to the free energy, which should match that of a single complex SYK model and have the symmetry $f(\mu)=f(-\mu)$. Expanding $\tilde{\lambda}_\alpha$ to the second-order gives
\begin{equation}
\begin{aligned}
\frac{S^{(1)}_{\text{eff}}}{N}
=&-\frac{\alpha_S}{J}\int d\tau \left[\text{Sch}\left(\tanh\frac{\varphi_L(\tau)}{2},\tau\right)+\text{Sch}\left(\tanh\frac{\varphi_R(\tau)}{2},\tau\right)\right]\\
&+\int d\tau \left[2f\left(\mu\right)+\frac{K}{2}\tilde{\lambda}_L'(\tau)^2+\frac{K}{2}\tilde{\lambda}_R'(\tau)^2\right].
\end{aligned}
\end{equation}
Here we define $K=-f''(\mu)=-f''(-\mu)$ and assume the winding number of $\tilde{\lambda}_\alpha$ is zero for the saddle points.

We then consider off-diagonal terms in $\sigma$, which is coupled to $G_{LR}$ and $G_{RL}$. Different from the diagonal terms where the IR fields are determined through expanding the quasi-solution around $G_{\infty,\mathcal{E}}(\tau_1-\tau_2)$, the off-diagonal components of $G$ it self can be considered as IR contributions. This can be understood from the fact that $G_{LR}$ in \eqref{conformal} can be viewed as introducing a real-time separation $\delta \tau =i \pi/t'$ to the $G_{LL}$. Consequently, we just add the contribution from $\mu_c$ directly as:
\begin{equation}
\frac{S^{(2)}_{\text{eff}}}{N}=-2b\mu_c\int d\tau\ \left(\frac{\varphi_R'(\tau)\varphi_L'(\tau)}{4\cosh^2 (\varphi_R(\tau)-\varphi_L(\tau)/2)}\right)^{1/4}\cos{\left(\tilde\lambda_R(\tau)-\tilde\lambda_L(\tau)\right)}.
\end{equation}
The total effective action is obtained by summing up diagonal and off-diagonal contributions $S_{\text{eff}}=S^{(1)}_{\text{eff}}+S^{(2)}_{\text{eff}}$.

We first consider the thermal equilibrium state. The conformal solution \eqref{conformal} corresponds to setting $\varphi_\alpha(\tau)=t'\tau$ and $\tilde{\lambda}_\alpha(\tau)=0$. This gives the grand potential $\Omega_{\text{WH}}$:
\begin{equation}
\frac{\Omega_{\text{WH}}}{N}=\text{Min}_{t'}\left\{2f(\mu)+\frac{\alpha_St'^2}{J}-\sqrt{2}b\mu_c (t')^{1/2}\right\}=2f(\mu)-\frac{3}{4}\left(\frac{Jb^4\mu_c^4}{\alpha_S}\right)^{1/3},
\end{equation}
with the solution $t'=\frac{b^{2/3} J^{2/3} \mu _c^{2/3}}{2 \alpha _S^{2/3}},$ where the $\mathcal{E}$ dependence comes form $b$. It is straightforward to check \eqref{conformal} then satisfies the saddle-point equations for $\varphi_\alpha$ and $\tilde{\lambda}_\alpha$, as well as the condition that all SL(2,R) charges should vanish \cite{maldacena2018eternal}. 

The charge of the left system is given by taking derivative with $\mu_L$. This gives $-Q_L=f'(\mu)$, meaning the relation between $\mu$ and $Q_L$ is the same as the thermodynamics of a single SYK at $T=0$. It is known that for a single SYK model at finite temperature, the chemical potential shifts as $\mu_s(0)=\mu_s(T)+2\pi\mathcal{E}_s/\beta$ \cite{gu2020notes}. This gives the relation between $\mu'$ and $\mu$ as
\begin{equation}\label{eqmu}
\mu=\mu'+2\pi\mathcal{E}/\beta_{\text{eff}}.
\end{equation}

We can also study the fluctuation of low-energy modes around the saddle point. The vanishing of SL(2,R) charges makes it possible to choose $\varphi_L(\tau)=\varphi_R(\tau)=\varphi(\tau)$. For convenience, we further introduce $\varphi'(\tau)=e^{\phi(\tau)}$ with the saddle point at $\phi(\tau)=\log(t')$. Expanding the effective action to the quadratic order gives
\begin{equation}
\begin{aligned}
\frac{\delta S_{\text{eff}}}{N}=\int d\tau \left[\frac{K}{2}\lambda_c'(\tau)^2+\frac{K}{2}\lambda_r'(\tau)^2+\sqrt{2}b\mu_ct'^{\frac{1}{2}}\lambda_r^2(\tau)+\frac{\alpha_S}{J}\delta\phi'(\tau)^2+\frac{3b^{4/3}\mu_c^{4/3}J^{1/3}}{8\alpha_S^{1/3}}\delta\phi(\tau)^2\right].
\end{aligned}
\end{equation}
Here we have defined $\lambda_c=\frac{\lambda_L+\lambda_R}{\sqrt{2}}$ and $\lambda_r=\frac{\lambda_L-\lambda_R}{\sqrt{2}}$. We find there is a gapped graviton mode with frequency $\omega_{\text{gravity}}= \sqrt{\frac{3}{8}} \left(\frac{b J \mu _c}{\alpha _S}\right)^{2/3}$, a relative phase mode $\lambda_r$ with frequency $\omega_{U(1)}=\frac{\sqrt{2} b^{2/3} J^{1/6} \mu _c^{2/3}}{\sqrt{K} (\alpha _S)^{1/6}}$, and a gapless field $\lambda_c$. These modes can be excited in certain quench dynamics, similar to the oscillation behavior observed in \cite{zhou2020disconnecting}. In addition to thermodynamics, the entanglement entropy of the wormhole phase \eqref{entropy} can also be computed using the effective action following the derivation in \cite{Chen:2019qqe}, since $\tilde{\lambda}_\alpha=0$ still satisfies the saddle-point equation after introducing replicas. 

For large $\bar{\mu}/t'$ and moderate $\mu/J$, the wormhole solution \eqref{conformal} breaks down, and the system transits into the black hole solution. The transition point is determined by comparing the free energy. In the black hole phase, we expect the free energy takes the form:
\begin{equation}
\frac{\Omega_{\text{BH}}}{N}=f(\mu_L)+f(\mu_R)+O(\mu_c^2)\approx f(\bar{\mu}+\mu)+f(\bar{\mu}-\mu)\approx 2f(\mu)-\bar{\mu}^2K.
\end{equation}
The critical $\bar{\mu}$ is then determined as $\text{Min}\left\{\frac{b^{2/3} J^{2/3} \mu _c^{2/3}}{8 \alpha _S^{2/3}},\frac{\sqrt{3} b^{2/3} {J}^{1/6} \mu _c^{2/3}}{2 \sqrt{K} {\alpha _S}^{1/6}}\right\}$, both of which proportional to $\mu_c^{2/3}$. This also justifies the neglect of higher order terms $\sim \mu_c^2$.

\subsection{Quench Dynamics}\label{quenchsec}
We still need to relate the TFD parameter $\beta_{\text{eff}}$ to parameters in the Hamiltonian. Following the idea in \cite{maldacena2018eternal}, this can be determined by considering the quench dynamics for turning off $\mu_c$. 

We firstly transform the effective action into the real-time with $\tau \rightarrow iu$ and $\varphi \rightarrow i\varphi$. Explicitly, we have
\begin{equation}
\begin{aligned}
\frac{\tilde{S}_{\text{eff}}}{N}=&\int du \left[-\frac{2\alpha_S}{J}\text{Sch}\left(\tan\frac{\varphi(u)}{2},u\right)+\frac{K}{2}\tilde{\lambda}_L'(u)^2+\frac{K}{2}\tilde{\lambda}_R'(u)^2\right]\\
&+\sqrt{2}b\int du\ \mu_c \theta(-u)\varphi'(u)^{1/2}\cos{\left(\tilde{\lambda}_R(u)-\tilde{\lambda}_L(u)\right)}.
\end{aligned}
\end{equation}
At $u<0$, the system is in thermal equilibrium with $\varphi(u)=t'u$ and $\tilde{\lambda}_\alpha(u)=0$. For $u>0$, by solving the saddle-point equations and requiring the smoothness for fields $\phi(u)=\log (\varphi(u)), \tilde{\lambda}_\alpha(u)$ and their first-order derivatives, we have for $u>0$
\begin{equation}
\varphi(u)=2\arctan\left(\tanh\frac{t'u}{2}\right),\ \ \ \ \ \ \ \ \ \ \ \ \tilde{\lambda}_\alpha(u)=0.
\end{equation}
We can derive the real-time Green's functions by analytical continuation of \eqref{IRqusi}. For example, 
\begin{equation}\label{G>q}
iG_{LL}^>(u_1,u_2)\equiv\left<c^{}_{L,i}(u_1)c^{\dagger}_{L,i}(u_2)\right>=b\times e^{\pi\mathcal{E}+i\mathcal{P}(u_1-u_2)}\left(\frac{\varphi'(u_1)\varphi'(u_2)}{4\sin^2 \left(i\frac{\varphi(u_1)-\varphi(u_2)}{2}+0^+\right)}\right)^{1/4}
\end{equation}
When $u_1>0$ and $u_2>0$, this gives
\begin{equation}\label{quencheff}
iG_{LL}^>(u_1,u_2)=be^{\pi\mathcal{E}+i\mathcal{P}u_{12}}\left(\frac{t'}{2\sinh \left(t' |u_{12}|/2\right)}\right)^{1/2}e^{-i\frac{\pi}{4}\text{sgn}(u_{12})}.
\end{equation}

We compare \eqref{quencheff} with the expectation if we assume the ground state is a TFD. The result is an analytical continuation of \eqref{quench}, which reads
\begin{equation}\label{quenchtfd}
\begin{aligned}
i\tilde{G}_{LL}(u)&=\left<G \right|e^{i H_0 u}c^{}_{L,i}e^{-i H_0 u}c^{\dagger}_{L,i}\left|G\right>=be^{\pi\mathcal{E}+i\mathcal{P}u_{12}}\left(\frac{\pi}{\beta_{\text{eff}}\sinh \left(\pi |u_{12}|/\beta_{\text{eff}}\right)}\right)^{1/2}e^{-i\frac{\pi}{4}\text{sgn}(u_{12})},
\end{aligned}
\end{equation}
where we have used \eqref{eqmu}. The equivalence between \eqref{quencheff} and \eqref{quenchtfd} gives $t'\beta_{\text{eff}}=2\pi$.

\section{Numerical Results} \label{III}
In this section, we present numerical results to support our low-energy analysis. In subsection \ref{im}, we solve the self-consistent equation in the imaginary time and compare the Green's function the conformal solutions \eqref{conformal}. We also discuss the existence of several phases for our model \eqref{Ham}. In subsection \ref{re}, we study the quench dynamics and comparing numerics with the prediction of the TFD state. 

\subsection{Imaginary time results and the Phase diagram} \label{im}
\begin{figure}[t]
  \center
  \includegraphics[width=0.95\columnwidth]{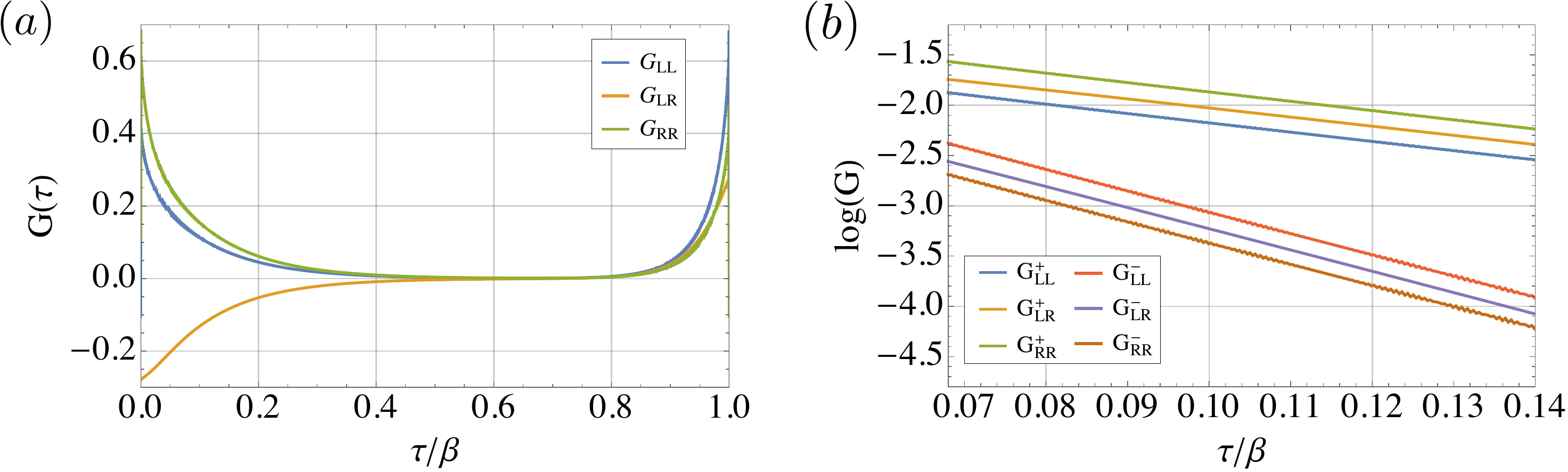}
  \caption{(a). The imaginary-time Green's function for $\beta J=120$, $\bar{\mu}/J=0.05$, $\mu/J=0.1$, and $\mu_c/J=0.05$. (b). A zoom in of (a) near $1/t'\ll\tau\ll\beta/2$ or $1/t'\ll\beta-\tau\ll\beta/2$. We can see the numerical results are consistent with \eqref{conformal}.} \label{fig1}  
 \end{figure}

\begin{figure}[t]
  \center
  \includegraphics[width=0.94\columnwidth]{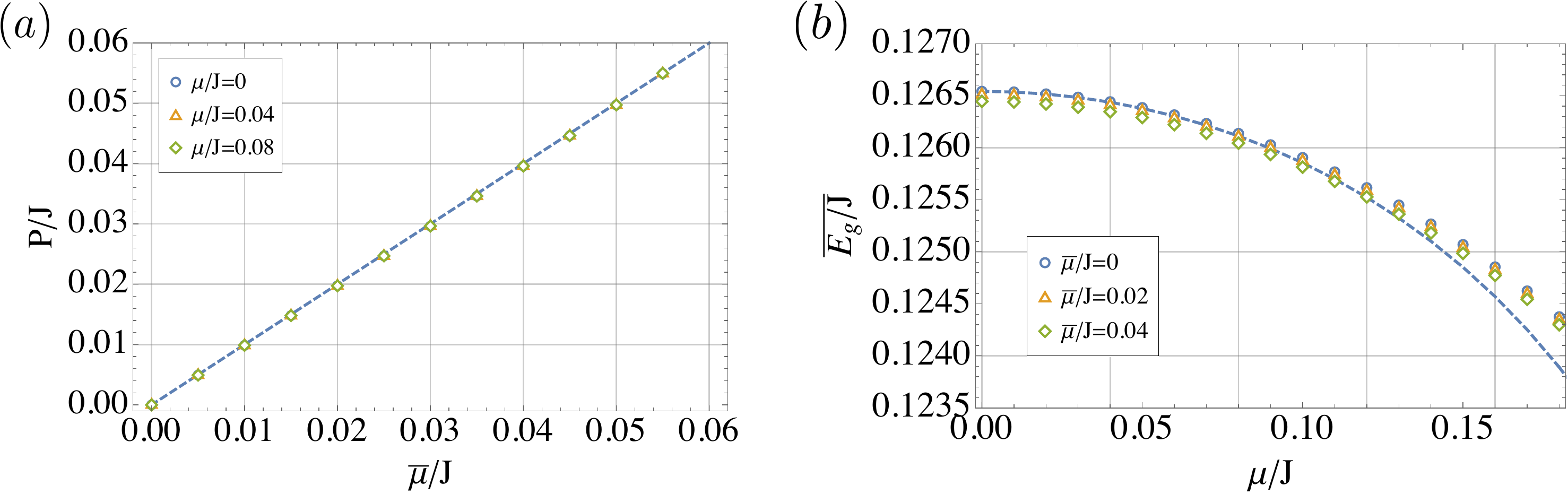}
  \caption{A comparison between the numerics and \eqref{compare} for (a). $\mathcal{P(\bar{\mu})}$ with different $\mu/J$, and (b). $\bar{E}_g(\mu)$ for different $\bar{\mu}$. In all cases we fix $\beta J=120$ and $\mu_c/J=0.05$. The dots are numerical data and the dashed line represents the prediction of \eqref{compare}. } \label{fig2}  
 \end{figure}

In the imaginary-time approach, we adapt the standard algorithm \cite{maldacena2018eternal} to solve the saddle-point equation \eqref{SD} for $G_{\alpha\beta}$ and $\Sigma_{\alpha\beta}$. We work at large but finite temperature $\beta J\gg1$ and small but finite $\mu_c/J$. 

Typical Green's functions in the wormhole phase are shown in Figure \ref{fig1} (a), where we choose $\beta J=120$, $\bar{\mu}/J=0.05$, $\mu/J=0.1$, and $\mu_c/J=0.05$. The Green's functions decay exponentially for $1/t'\ll\tau\ll\beta/2$ or $1/t'\ll\beta-\tau\ll\beta/2$. We can approximately identify the numerical solution at $0<\tau<\beta/2$ with the zero-temperature Green's function $G^+_{\alpha\beta}(\tau)\equiv |G_{\alpha\beta}(\tau)|$, and the numerical solution at $0<\beta-\tau<\beta/2$ with the zero-temperature Green's function $G^-_{\alpha\beta}(\tau)\equiv |G_{\alpha\beta}(\tau-\beta)|$. For $1/t'\ll\tau\ll\beta/2$, the conformal solution predicts
\begin{equation}
\begin{aligned}
&\log G^{+}_{LL}(\tau) \sim c_0+\pi\mathcal{E}-(\bar{E}_g-\mathcal{P})\tau,\ \ \ \ \ \ \ \ \log G^{-}_{LL}(\tau) \sim c_0-\pi\mathcal{E}-(\bar{E}_g+\mathcal{P})\tau, \\
&\log G^{+}_{RR}(\tau) \sim c_0-\pi\mathcal{E}-(\bar{E}_g-\mathcal{P})\tau,\ \ \ \ \ \ \ \ \log G^{-}_{RR}(\tau) \sim c_0+\pi\mathcal{E}-(\bar{E}_g+\mathcal{P})\tau, \\
&\log G^{+}_{LR}(\tau) \sim c_0-(\bar{E}_g-\mathcal{P})\tau,\ \ \ \ \ \ \ \ \ \ \ \ \ \ \ \ \log G^{-}_{LR}(\tau) \sim c_0-(\bar{E}_g+\mathcal{P})\tau,
\end{aligned}
\end{equation}
where $c_0$ is come numerical factor and we have defined the averaged gap $\bar{E}_g=t'/4$. In Figure \ref{fig1} (b), we show the numerical results, which takes exactly the same form as conformal solutions. 

\begin{figure}[t]
  \center
  \includegraphics[width=0.95\columnwidth]{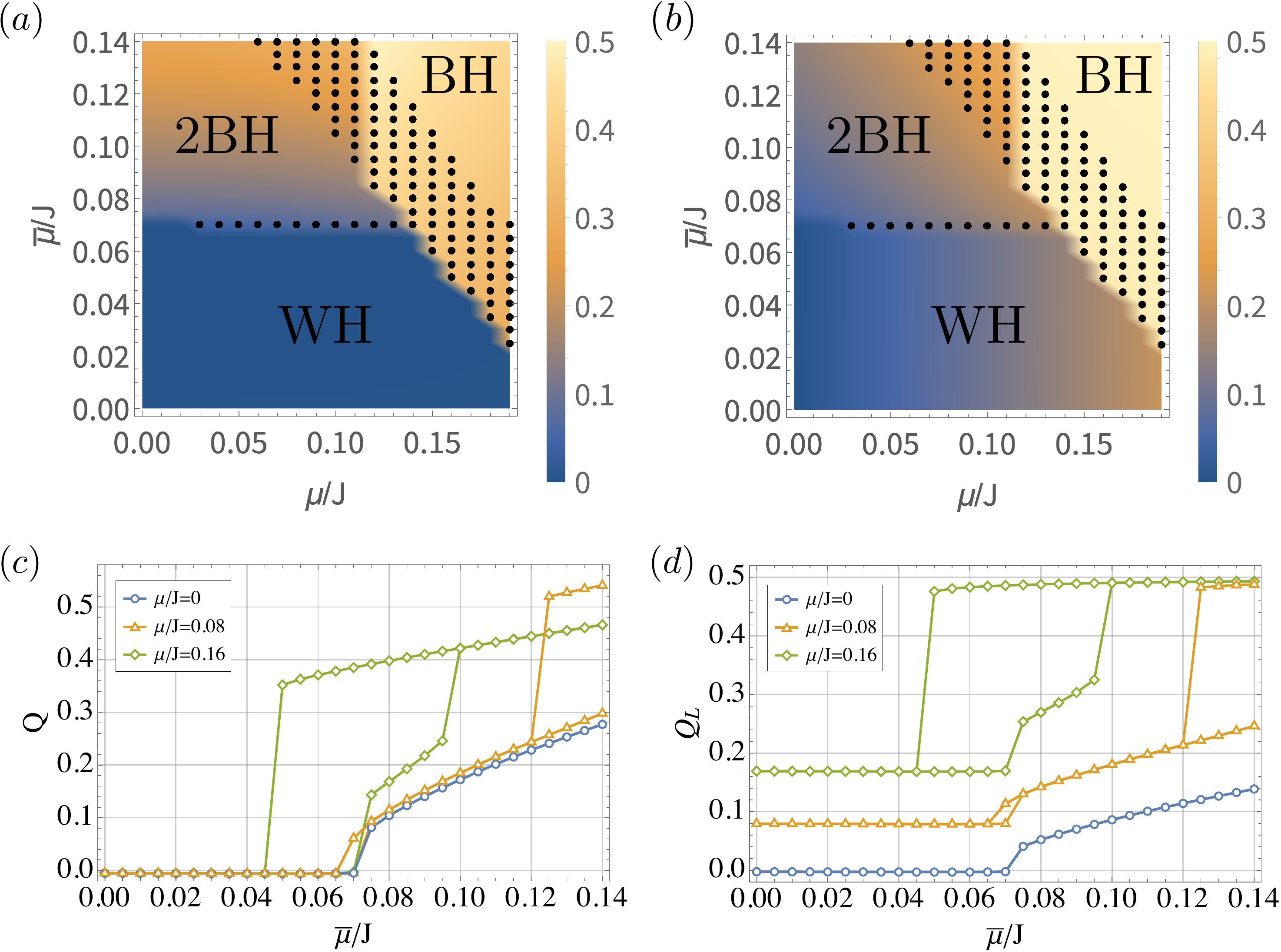}
  \caption{The phase diagram of the model considered in this paper with $\beta J=120$ and $\mu_c/J=0.05$. Here we focus on moderate $\bar{\mu}/J$, with the wormhole phase, the two black hole phase, and the single black hole phase. We show the density plot for $Q$ in (a) and for $Q_L$ in (b). Multiply saddle points exist in the dotted region. We also plot corresponding 1D curves with fixed $\mu/J$ in (c) and (d). } \label{fig3}  
 \end{figure}

We then compare the numerical fitted $\mathcal{P}$ and $\bar{E}_g$ with the low-energy prediction summarized in Table \ref{tab}, where
\begin{equation}\label{compare}
\mathcal{P}=\bar{\mu},\ \ \ \ \ \ \ \ \ \ \bar{E}_g(\mu)=\bar{E}_g(0)/\cosh(2\pi\mathcal{E})^{1/6}.
\end{equation} 
The result is shown in Figure \ref{fig2} for $\beta J=120$ and $\mu_c/J=0.05$. We find the results match well, with a small discrepancy possibly from the fact that $\mu_c/J$ is not small enough.

We then consider large $\bar{\mu}$ and explore the phase diagram for the Hamiltonian \eqref{Ham}. For fixed $\mu$, we perform the numerical iteration firstly from small $\bar{\mu}$ to large $\bar{\mu}$, and then from large $\bar{\mu}$ back to small $\bar{\mu}$. We also perform similar steps for different $\mu$ with fixed $\bar{\mu}$. Near first-order transition points, this can give different saddle points for the same $\bar{\mu}$. 

The results for $\beta J=120$ and $\mu_c/J=0.05$ are shown in Figure \ref{fig3}. At small $\bar{\mu}/J$, the system is in the wormhole phase, with specific charge $Q=0$, while $Q_L$ can change when tunning $\mu/J$. For moderate $\mu/J$, if we consider larger $\bar{\mu}/J$, the system transits into the two black hole phase. In this phase, we can approximate the solution as two separate complex SYK models with chemical potential $\mu_L$ and $\mu_R$. In this case, both $Q_L$ and $Q$ change freely when tunning parameters. If we further consider a larger $\mu$ or $\bar{\mu}$, where $\mu_L$ becomes larger than a threshold $\sim 0.25$, it is known that a single complex SYK model would transit into a (nearly) fully occupied phase with $Q_L=1/2$ \cite{azeyanagi2018phase,patel2019theory,tikhanovskaya2020excitation}. We call this phase the (single) black hole phase since the right system can be approximated as a single complex SYK black hole. One also expects a (nearly) fully occupied phase for both copies to exist when both $\mu_L$ and $\mu_R$ become large and the system has $Q_L=Q_R=1/2$.

\subsection{Real time results and Quench dynamics}\label{re}
Now we turn to real-time formalism. For the Majorana coupled SYK model, related studies have been carried out in \cite{Qi_2020,Plugge_2020,maldacena2019syk}. We are mainly interested in study the quench dynamics, and compare the numerical to results in \ref{quenchsec}. 

\begin{figure}[t]
  \center
  \includegraphics[width=0.95\columnwidth]{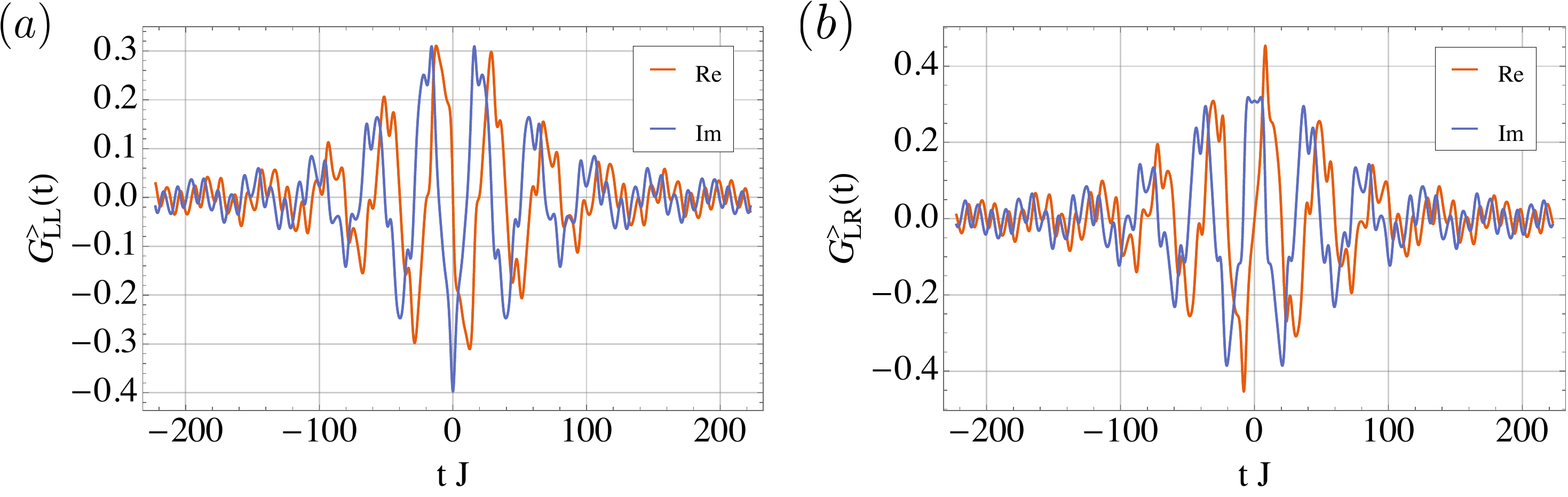}
  \caption{The real-time Green's functions $G^>(t)$ in thermal equilibrium within the wormhole phase. We choose $\beta J=40$, $\mu_c/J=0.075$, $\mu/J=0.1$ and $\bar \mu/J=0$. } \label{fig4}  
 \end{figure}

The quench dynamics of SYK-like models can be studied using the Keldysh approach \cite{eberlein2017quantum,zhang2019evaporation,almheiri2019universal,cheipesh2020quantum,kuhlenkamp2020periodically}. We firstly define several real-time Green's functions $G^>$ and $G^<$:
\begin{equation}
G^>_{\alpha\beta}(u_1,u_2)=-i\left<c^{}_{\alpha,i}(u_1)c^{\dagger}_{\beta,i}(u_2)\right>,\ \ \ \ \ \ \ \ \ \  G^<_{\alpha\beta}(u_1,u_2)=i\left<c^{\dagger}_{\beta,i}(u_2)c^{}_{\alpha,i}(u_1)\right>.
\end{equation}
Here we have copied the definition \eqref{G>q} for completeness. The retarded and advanced Green's functions $G^R$ and $G^A$ then read
\begin{equation}\label{GRA}
\begin{aligned}
&G^R_{\alpha\beta}(u_1,u_2)=\theta(u_{12})\left(G^>_{\alpha\beta}(u_1,u_2)-G^<_{\alpha\beta}(u_1,u_2)\right),\\
&G^A_{\alpha\beta}(u_1,u_2)=\theta(-u_{12})\left(G^<_{\alpha\beta}(u_1,u_2)-G^>_{\alpha\beta}(u_1,u_2)\right).
\end{aligned}
\end{equation}
For the Hamiltonian \eqref{Ham}, the self-energy $\Sigma^>$ and $\Sigma^<$ in real-time can be obtained by the analytically continuation of \eqref{SD} in the time domain:
\begin{equation}\label{Sreal}
\begin{aligned}
&\Sigma^>_{\alpha\beta}(u_1,u_2)=J^2G^>_{\alpha\beta}(u_1,u_2)^2G^<_{\beta\alpha}(u_2,u_1),\\
&\Sigma^<_{\alpha\beta}(u_1,u_2)=J^2G^<_{\alpha\beta}(u_1,u_2)^2G^>_{\beta\alpha}(u_2,u_1),
\end{aligned}
\end{equation}
and the retarded and advanced components of the self-energy is 
\begin{equation}\label{SRA}
\begin{aligned}
&\Sigma^R_{\alpha\beta}(u_1,u_2)=\theta(u_{12})\left(\Sigma^>_{\alpha\beta}(u_1,u_2)-\Sigma^<_{\alpha\beta}(u_1,u_2)\right),\\
&\Sigma^A_{\alpha\beta}(u_1,u_2)=\theta(-u_{12})\left(\Sigma^<_{\alpha\beta}(u_1,u_2)-\Sigma^>_{\alpha\beta}(u_1,u_2)\right).
\end{aligned}
\end{equation}

\begin{figure}[t]
  \center
  \includegraphics[width=0.95\columnwidth]{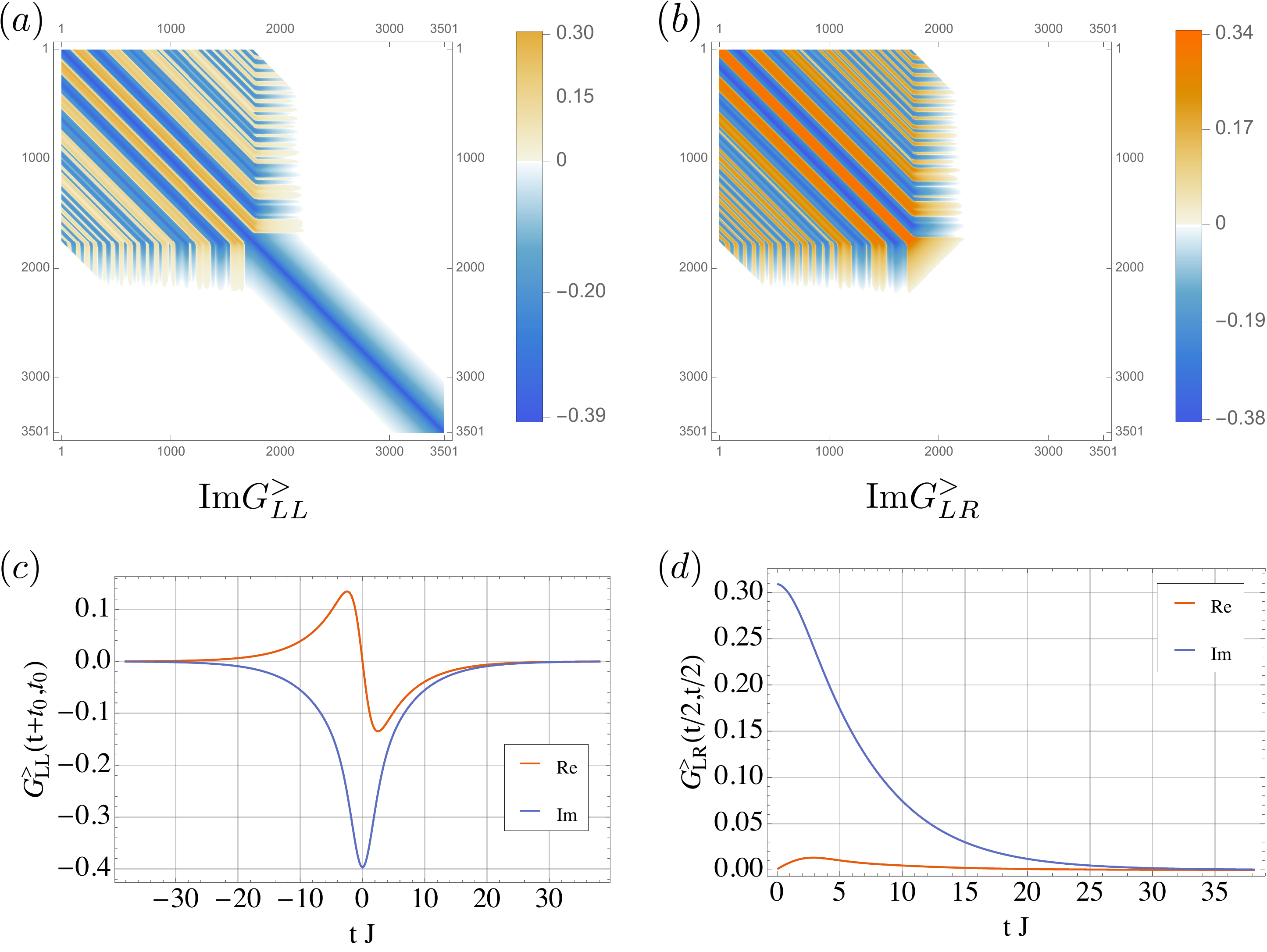}
  \caption{The quench dynamics when turning off $\mu_c$ with $\beta J=40$, $\mu_c/J=0.075$ and $\bar \mu/J=0$. We plot Im$G^>_{LL}(u_1,u_2)$ in (a) and Im$G^>_{LR}(u_1,u_2)$ in (b). When $u_1>0$ and $u_2>0$, we find $G^>_{LL}(u_1,u_2)$ only depends on $u_1-u_2$, while $G^>_{LR}(u_1,u_2)$ only depends on $u_1+u_2$. We also plot $G^>_{LL}(t+t_0,t_0)$ and $G^>_{LR}(t/2,t/2)$ explicitly in (c) and (d). } \label{fig5}  
 \end{figure}
Defining $\hat{\mu}(u_1)=\bar\mu I +\mu \sigma_z+\mu_c(u_1) \sigma_x$, we have matrix equations
\begin{equation}\label{SDGR}
\begin{aligned}
i\partial_{u_1}G^{R/A}(u_1,u_2)+\hat{\mu}(u_1)G^{R/A}(u_1,u_2)-\int du_3\Sigma^{R/A}(u_1,u_3)G^{R/A}(u_3,u_2)&=\delta({u_{12}})I,\\
-i\partial_{u_2}G^{R/A}(u_1,u_2)+G^{R/A}(u_1,u_2)\hat{\mu}(u_2)-\int du_3G^{R/A}(u_1,u_3)\Sigma^{R/A}(u_3,u_2)&=\delta({u_{12}})I.
\end{aligned}
\end{equation}
\begin{figure}[t]
  \center
  \includegraphics[width=0.9\columnwidth]{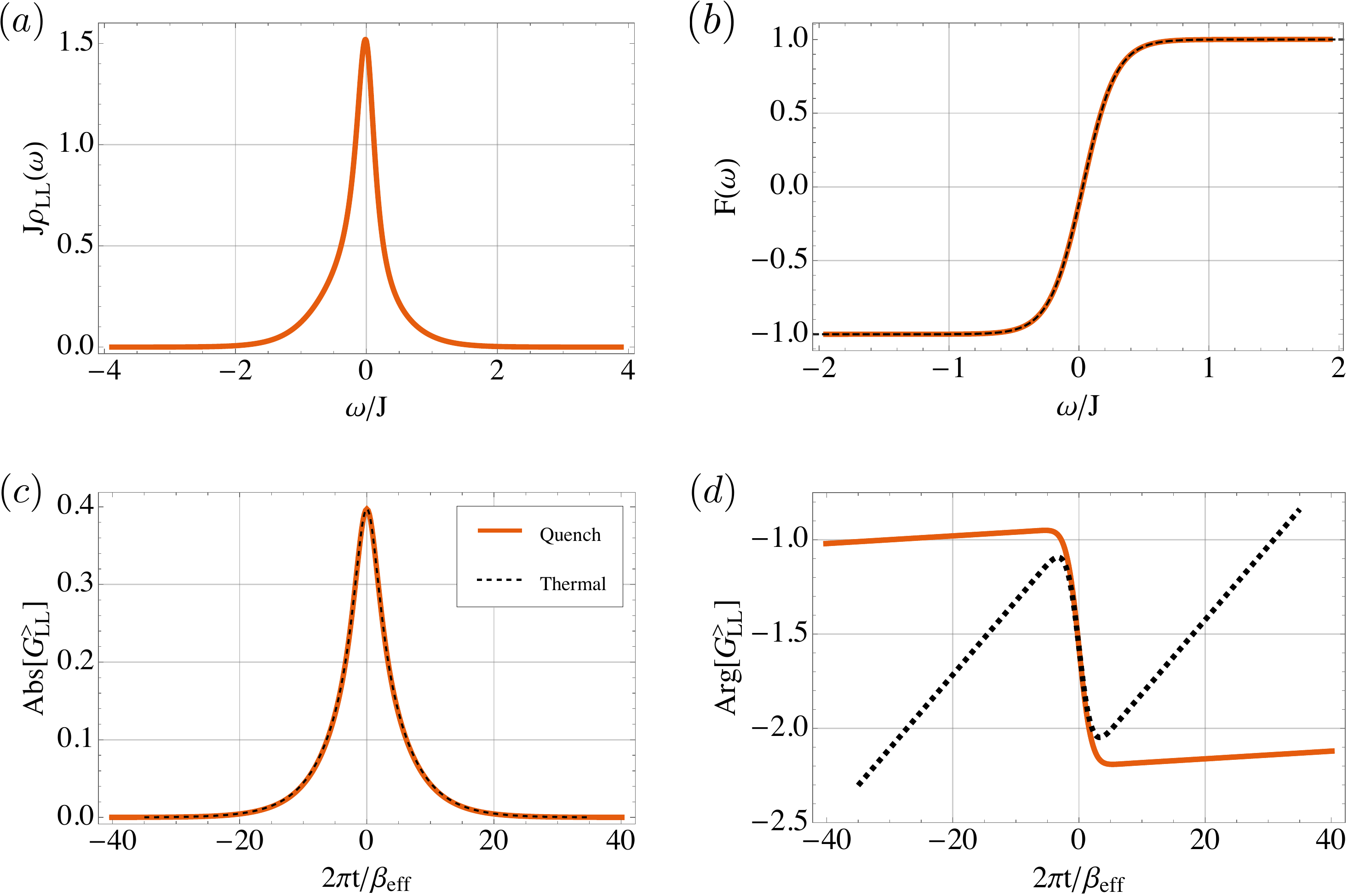}
  \caption{(a). The spectral function $\rho_{LL}(\omega)$ in the long time limit for turning off $\mu_c$. (b). The quantum distribution $F(\omega)$ in the long time limit for turning off $\mu_c$. The dashed line represents the fitting with $F(\omega)=\tanh\left(3.95 (\omega/J-0.029)\right)$, which gives a final state with $\beta_\text{eff} J=7.9$ and $\mu'/J=0.129$. We then compare the thermal real-time Green's function and the quench result in (c) and (d), with only a phase difference due to $\delta \mu$, as predicted by \eqref{quenchtfd}.} \label{fig6}  
 \end{figure}

To study the quench protocol in section \ref{quenchsec}, we should prepare a system at thermal equilibrium in the wormhole phase. All Green's functions and self-energies in thermal ensemble only depend on the time difference. The inverse temperature $\beta$ enters the self-consistent equation through the fluctuation-dissipation theorem:
\begin{equation}\label{FDT}
G^>_{\alpha\beta}(\omega)=-i\rho_{\alpha\beta}(\omega)n_F(-\omega,\beta),\ \ \ \ \ \ \ \ \ G^<_{\alpha\beta}(\omega)=i\rho_{\alpha\beta}(\omega)n_F(\omega,\beta).
\end{equation}
with the spectral function matrix $\rho_{\alpha\beta}(\omega)$ defined as
\begin{equation}\label{rho}
\rho_{\alpha\beta}(\omega)=-\frac{1}{2\pi i}\left(G^R_{\alpha\beta}(\omega)-G^A_{\alpha\beta}(\omega)\right)=-\frac{1}{2\pi i}\left(G^R_{\alpha\beta}(\omega)-G^R_{\beta\alpha}(\omega)^*\right).
\end{equation}
Solving \eqref{Sreal}, \eqref{SRA}, \eqref{SDGR}, \eqref{FDT} and \eqref{rho} iteratively with $\mu_c(u_1)=\mu_c$ gives the spectral function and real-time Green's functions. In the numerics, we choose $\beta J=40$, $\mu_c/J=0.075$, $\mu/J=0.1$ and $\bar \mu/J=0$. The results for $G^>_{LL}$ and $G^>_{LR}$ are shown in Figure \ref{fig4} as an example, which show rapid oscillations as in \cite{Plugge_2020,Qi_2020}.

Now we consider the quench dynamics with $\mu_c(u)=\mu_c \theta(-u)$. The evolution of Green's functions is usually determined using the Kadanoff-Baym equation, which takes a similar form as \eqref{SDGR}:
\begin{equation}\label{KBequation}
\begin{aligned}
i\partial_{1}G^{\gtrless}(u_1,u_2)+\hat{\mu}(u_1)G^{\gtrless}(u_1,u_2)&=\int du_3\left(\Sigma^R(u_1,u_3)G^{\gtrless}(u_3,u_2)+\Sigma^\gtrless(u_1,u_3)G^{A}(u_3,u_2)\right),\\
-i\partial_{2}G^{\gtrless}(u_1,u_2)+G^{\gtrless}(u_1,u_2)\hat{\mu}(u_2)&=\int du_3\left(G^R(u_1,u_3)\Sigma^{\gtrless}(u_3,u_2)+G^\gtrless(u_1,u_3)\Sigma^{A}(u_3,u_2)\right).
\end{aligned}
\end{equation}
Importantly, the causality is preserved explicitly due to the presence of retarded and advanced components on the R.H.S., and the derivative of the Green's functions at time $(u_1,u_2)$ only depends on Green's functions $G^{\gtrless}(u_3,u_4)$ with $u_3,u_4<\text{Max}\{u_1,u_2\}$. The initial condition for \eqref{KBequation} is given by the thermal equilibrium result
\begin{equation}
G^{\gtrless}(u_1,u_2)=G^{\gtrless}(u_1-u_2), \ \ \ \ \ \ \ \text{for $u_1<0$ and $u_2<0$.}
\end{equation}
We then evolve the Green's function step by step using the standard numerical method by discretizing $G^{\gtrless}(u_1,u_2)$ into large matrices \cite{eberlein2017quantum,zhang2019evaporation,almheiri2019universal,cheipesh2020quantum,kuhlenkamp2020periodically}. 

The numerical results for $G^{>}(u_1,u_2)$ are shown in Figure \ref{fig5}. As we turn off the coupling between two systems, their correlation $G_{LR}$ decays exponentially. More interestingly, we see that $G^>_{LL}(u_1,u_2)$ only depends on $u_1-u_2$, while $G^>_{LR}(u_1,u_2)$ only depends on $u_1+u_2$ when $u_1>0$ and $u_2>0$. This is consistent with what one expects for the evolution of the TFD state.

To compare the prediction from the TFD perspective and numerical results more carefully, we need to extract the temperature and the chemical potential of the final state. We use the numerical results $G^{\gtrless}(u_1-u_2)$ after the quench in the long-time limit. We show the long-time spectral function $\rho_{LL}$ in Figure \ref{fig6} (a). The fluctuation-dissipation theorem predicts
\begin{equation}
-\frac{G^{>}_{LL}(\omega)+G^{<}_{LL}(\omega)}{G^{>}_{LL}(\omega)-G^{<}_{LL}(\omega)}=\tanh\frac{\beta_{\text{eff}}(\omega+\delta\mu)}{2}.
\end{equation}
We then determine $\beta_{\text{eff}}$ and $\delta\mu$ by fitting, as shown in Figure \ref{fig6} (b). This gives $\beta_\text{eff} J=7.9$ and $\mu'/J=0.129$. 

Using this result, we can perform an independent numerical study by solving a single model at corresponding parameters. In Figure \ref{fig6} (c-d), we compare the single SYK numerics with the result of quench dynamics. We find the magnitude of $G_{LL}$ matches to high accuracy, while a phase difference exists due to the presence of $\delta \mu$. As expected from \eqref{quenchtfd}, in the quench dynamics, the phase is almost constant in the long-time limit, while approaching $-\pi/4$ or $-3\pi/4$ in the short-time limit.

\section{Summary} \label{IIII}
We study the generalization of the coupled SYK model \eqref{Ham} by introducing $U(1)$ symmetry. We find the conformal solution \eqref{conformal} at $T=0$, which contains IR parameters $\mathcal{E}$, $\mathcal{P}$ and $t'$. We determine the IR parameter, as shown in Table \ref{tab}, by combining understandings from the TFD and the effective action analysis. We also determine the TFD parameters as shown in Table \ref{tab2}. We then compare our results with numerics and explore the full phase diagram.

A few possible extensions of the current work are as follows: One can ask whether we could write out a gravity theory with the same effective action $S_{\text{eff}}$. This should correspond to adding $U(1)$ gauge fields \cite{sachdev2019universal} to the JT gravity with coupling between two boundaries \cite{maldacena2018eternal}.  It is also interesting to consider two copies of higher-dimensional models, etc. 1+1-D, and adding a coupling term. This can lead to higher-dimensional wormholes. 

\section*{Acknowledgments}

We especially thank Yingfei Gu for helpful discussions.
P.Z. 
acknowledges support from the Walter Burke Institute for Theoretical Physics at Caltech.
\bibliographystyle{JHEP}
\bibliography{ref.bib}

\end{document}